\documentclass[12pt]{amsart}

\usepackage{amsmath,amssymb,amsbsy,amsfonts,amsthm,latexsym,
amsopn,amstext,amsxtra,euscript,amscd,mathrsfs,color,bm,cite}

\usepackage{float} 
\usepackage[english]{babel}
\usepackage{mathtools}
\usepackage{todonotes}
\usepackage{url}
 \usepackage{soul}
\usepackage[colorlinks,linkcolor=blue,anchorcolor=blue,citecolor=blue,backref=page]{hyperref}
\RequirePackage{mathrsfs} \let\mathcal\mathscr

\usepackage{enumitem}

\def\le{\leqslant}
\def\ge{\geqslant}

\def\dgP{\deg P} 

\def\leq{\leqslant}
\def\geq{\geqslant}

\usepackage{mathtools}
\usepackage{todonotes}
\usepackage[norefs,nocites]{refcheck}

\usepackage{amsmath,amssymb,latexsym,rotating}

\begin{document}

\newtheorem{theorem}{Theorem}
\newtheorem{corollary}[theorem]{Corollary}
\newtheorem{definition}[theorem]{Definition}
\newtheorem{eg}[theorem]{Example}
\newtheorem{lemma}[theorem]{Lemma}
\newtheorem{remark}[theorem]{Remark}
\newtheorem{proposition}[theorem]{Proposition}

\numberwithin{equation}{section}
\numberwithin{theorem}{section}
\numberwithin{table}{section}
\numberwithin{figure}{section}

%
%



\def \balpha{\bm{\alpha}}
\def \bbeta{\bm{\beta}}
\def \bgamma{\bm{\gamma}}
\def \bdelta{\bm{\delta}}
\def \blambda{\bm{\lambda}}
\def \bmu{\bm{\mu}}
\def \bchi{\bm{\chi}}
\def \bphi{\bm{\varphi}}
\def \bpsi{\bm{\psi}}
\def \bpi{\bm{\pi}}
\def \bomega{\bm{\omega}}
\def \btheta{\bm{\vartheta}}

\def \bzeta{\bm{\zeta}}
\def \bxi{\bm{\xi}}

\def\a{\alpha}

\def\filledbox{\vrule height 1.8ex width 1.4ex depth 0ex }

\def\squareforqed{$\qed$}

\def\QED{\ifmmode\squareforqed\else{\unskip\nobreak\hfil
\penalty50\hskip1em\null\nobreak\hfil$\Box$
\parfillskip=0pt\finalhyphendemerits=0\endgraf}\fi}

\def\lcm{{\mathrm{lcm}}}

\def\({\left(}
\def\){\right)}
\def\fl#1{\left\lfloor#1\right\rfloor}
\def\rf#1{\left\lceil#1\right\rceil}

\def\mand{\qquad \mbox{and} \qquad}

\newcommand{\D}{D}
\newcommand{\Lcomp}{C}
\newcommand{\N}{N}
\newcommand{\w}{w}
\renewcommand{\k}{k}
\newcommand{\q}{q}
\renewcommand{\r}{r}

\newcommand{\ba}{\overline{a}}
\newcommand{\bb}{\mathbf{b}}
\newcommand{\bc}{\mathbf{c}}

\newcommand{\dm}{d_{\max}}
\newcommand{\Hm}{H_{\max}}

\newcommand{\bsm}[1]{\mbox{\boldmath \scriptsize$#1$}}
 
\newcommand{\iscript}{\bsm {i}}
 
\newcommand{\jscript}{\bsm {j}}

\newcommand{\bzero}{\mbox{\boldmath$0$}}
\newcommand{\bunit}{\mbox{\boldmath$I$}}
\newcommand{\bL}{\mbox{\boldmath$L$}}
\newcommand{\be}{\mbox{\boldmath$e$}}
 \newcommand{\bi}{\mbox{\boldmath$i$}}
 \newcommand{\bj}{\mbox{\boldmath$j$}}
 \newcommand{\bk}{\mbox{\boldmath$k$}}
\newcommand{\bv}{\mbox{\boldmath$v$}}
\newcommand{\bx}{\mbox{\boldmath$x$}}
\newcommand{\bu}{\mbox{\boldmath$u$}}
\newcommand{\bX}{\mbox{\boldmath$X$}}
\newcommand{\bZ}{\mbox{\boldmath$Z$}}

\newcommand{\eh}{\mbox{\em height}}

\newcommand{\h}{\mbox{height}}

\renewcommand{\rho}{r}

\newcommand{\cA}{{\mathcal A}}

\newcommand{\C}{{\mathbb C}}
\newcommand{\F}{{\mathbb F}}
\newcommand{\Q}{{\mathbb Q}}
\newcommand{\R}{{\mathbb R}}
\newcommand{\Z}{{\mathbb Z}}

\def \fE{E} 

\def \fA{\mathfrak A}

\def \fQ{\mathfrak Q}

\def \fh{{\mathfrak h}_a} 

\def\cA{{\mathcal A}}
\def\cB{{\mathcal B}}
\def\cC{{\mathcal C}}
\def\cD{{\mathcal D}}
\def\cE{{\mathcal E}}
\def\cF{{\mathcal F}}
\def\cG{{\mathcal G}}
\def\cH{{\mathcal H}}
\def\cI{{\mathcal I}}
\def\cJ{{\mathcal J}}
\def\cK{{\mathcal K}}
\def\cL{{\mathcal L}}
\def\cM{{\mathcal M}}
\def\cN{{\mathcal N}}
\def\cO{{\mathcal O}}
\def\cP{{\mathcal P}}
\def\cQ{{\mathcal Q}}
\def\cR{{\mathcal R}}
\def\cS{{\mathcal S}}
\def\cT{{\mathcal T}}
\def\cU{{\mathcal U}}
\def\cV{{\mathcal V}}
\def\cW{{\mathcal W}}
\def\cX{{\mathcal X}}
\def\cY{{\mathcal Y}}
\def\cZ{{\mathcal Z}}

\newcommand{\ignore}[1] { }

\title[Sets of linear forms of
maximal complexity]{On sets of linear forms of
\\
maximal complexity}

\author[M. Kaminski, I. E. Shparlinski and M. Waldschmidt]{\hspace{-7.5pt} Michael~Kaminski,~Igor~E.~Shparlinski~and~Michel~Waldschmidt}
\thanks{During this work, I.E.S.   was partially  supported   by ARC Grant~DP200100355.}
\address{Department of Computer Science,
Technion -- Israel Institute of Technology,
Haifa 3200003,
Israel}
\email{kaminski@cs.technion.ac.il}

\address{School of Mathematics and Statistics, University of New South Wales, Sydney, NSW 2052, Australia}
\email{igor.shparlinski@unsw.edu.au}

\address{Sorbonne Universit\'e, CNRS, IMJ-PRG, F-75005 Paris, France}
\email{michel.waldschmidt@imj-prg.fr}

\date{\today}

\begin{abstract}
We present a uniform description of sets of $m$ linear forms
in $n$ variables over the field of rational numbers whose computation
requires $m(n - 1)$ additions.
\end{abstract}



\maketitle
%

\thispagestyle{empty}

\section{Introduction}

\subsection{Motivation and background} 
Evaluating a set of a linear forms is a natural computation task
that frequently appears in both theory and applications.
For a matrix
\begin{equation}
\label{eq: matrix}
\Delta
= \begin{pmatrix}  
\delta_{1,1} & \delta_{1,2} & \cdots & \delta_{1,n}
\\
\delta_{2,1} & \delta_{2,2} & \cdots & \delta_{2,n}
\\
 \vdots & \vdots & \ddots   & \vdots
\\
\delta_{m,1} & \delta_{m,2} & \cdots & \delta_{m,n}
\end{pmatrix}
\end{equation}
and a column vector 
\begin{equation}
\label{eq: vector} 
\bx = (x_1, \ldots, x_n)^T 
\end{equation}  
linear forms are presented as a matrix-vector product
\begin{equation}
\begin{split} 
\label{eq: sets}
\begin{pmatrix}
\delta_{1,1} & \delta_{1,2} & \cdots & \delta_{1,n}
\\
\delta_{2,1} & \delta_{2,2} & \cdots & \delta_{2,n}
\\
 \vdots & \vdots & \ddots & \vdots
\\
\delta_{m,1} & \delta_{m,2} & \cdots & \delta_{m,n}
\end{pmatrix}
\begin{pmatrix}
x_1
\\
x_2
\\
\vdots
\\
x_n 
\end{pmatrix} &
=
\Delta\, \bx \\
&= \(\delta_{s,1} x_1 +  \cdots +  \delta_{s,n} x_n\)_{s=1}^m, 
\end{split} 
\end{equation}
in which
the matrix entries $\delta_{s,t}$ are fixed values and
the vector entries $x_s$ are varying inputs and
computations are by means of {\em linear} algorithms.
As expected, the complexity of a linear algorithm is
its number of additions and
we are interested in sets of linear forms of high complexity.

We denote the additive complexity  of~\eqref{eq: sets}, that is, the minimum complexity of a linear algorithm that computes $\Delta\, \bx$,
by $\cC(\Delta)$ and call it the {\em complexity} of $\Delta$.

Obviously, the set of linear forms~\eqref{eq: sets}
can be computed in $m(n - 1)$ additions.
However, in finite fields,
this trivial upper bound is not the best possible. 
By~\cite[Theorem~1]{Savage74}, over a finite field of $q$ elements, 
it can be computed in $O(mn / \log_q m)$ additions, where the implied constants are absolute.  
On the other hand (in finite fields), when $m = O(n)$,
there exist $\delta_{s,t}$, $s = 1,\ldots,m$ and $t = 1,\ldots,n$,
for which any computation
of~\eqref{eq: sets} requires $\Omega(mn / \log_q m)$ additions,
see~\cite[Section~5]{Savage74}.
In fact,
this lower bound holds for almost all $m \times n$ matrices
with $m = O(n)$,
see~\cite[Appendix~B.2]{KaminskiS21} for the precise statement and
the proof by a {\em counting argument}. 
Thus,
for each pair of positive integers $m$ and $n$ such that $m = O(n)$,
the entries of such a matrix can be   computed by brutal force, namely by an exhaustive search, 
However,  to describe them explicitly or, at least, uniformly (in $m$ and $n$) is
a very difficult open problem.
Even no example of a non-linear complexity is
known from the literature.

The situation is quite different when the underlying field is infinite.
By a {\em transcendence degree argument},
it is easy to see that, over the field of real numbers $\R$, say,
when the entries of $\Delta$ are algebraically independent,
the computation of~\eqref{eq: sets} requires $m(n - 1)$ additions
(see~\cite[Section~5.2]{BuergisserCSL97}).
This leads to a natural question:
what about the field of rational numbers~$\Q$?
As it has been remarked in~\cite[Appendix~B.3]{KaminskiS21},
almost all matrices~\eqref{eq: matrix} are of such complexity and
a specific example of such a matrix is the main result
of~\cite{KaminskiS21}.  We also show in Theorem~\ref{thm: non-eff} that there 
is   such a matrix with reasonably small integer entries. However, 
our proof of this result  is not constructive, see Appendix~\ref{sec:count},  
and to find such an example
one  may have to search through all matrices with entries of the size described in Theorem~\ref{thm: non-eff}.

It has been shown in~\cite{KaminskiS21} that,
if the entries of a matrix $\Delta$ are algebraically independent and
$\Gamma$ is ``sufficiently close'' to $\Delta$,
then also $\cC(\Gamma) = m (n - 1)$.
However, an estimate of the above ``sufficiently close'' and,
as a corollary, a uniform description of such matrices $\Gamma$
is based on very non-trivial number-theoretic tools~\cite{Sert99} and
also involves lengthy and somewhat tedious calculations.

In a few words, the construction of $\Gamma$ in~\cite{KaminskiS21},
that strongly resembles the  pioneering work of  Strassen~\cite{Strassen74} 
of  1974, consists of four stages and is as follows.

First, by following the proof for the algebraically independent case,
it is shown that matrices $\Delta$ with $\cC(\Delta) < m(n - 1)$
are defined by polynomials possessing a rather simple structure and
a theorem of Perron~\cite{Perron27} is used to bound
the polynomials' degree and height.
Using these polynomials,
it is shown that if $\Delta$ is a real matrix whose entries are
algebraically independent (implying $\cC(\Delta) = m(n - 1)$),
and
$\Gamma$ is a matrix over $\Q$ that is sufficiently close to $\Delta$
(in the Frobenius norm), then $\cC(\Gamma) = m(n - 1)$ as well.
The rest is to construct such $\Delta$ and $\Gamma$.

The construction of $\Delta$ uses an effective version of
the Lindemann--Weierstrass theorem in
transcendental number theory due to Sert~\cite{Sert99}. 
In particular,
the entries of $\Delta$ are real numbers of the form $e^\alpha$,
where $\alpha = 2^{i/mn}$, $i = 1,\ldots,mn$.

Then, the construction of $\Gamma$ that is sufficiently close to $\Delta$,
is by an appropriate truncation of the Taylor expansions of $e^\alpha$.

Finally, $\Gamma$ is converted into an $m \times n$ integer matrix
$\Omega$ of complexity $m(n - 1)$.
Because of the approximation precision required by Sert's theorem,
the entries of $\Omega$ are triple exponential in the matrix size,
implying that their, say, binary representation is
double exponential in $mn$.

To the best of our knowledge, this is the only example of a set
of linear forms over $\Q$ of a non-linear complexity.  In fact, this set is 
of the largest possible complexity.

\subsection{New construction} 
In this paper we present an example of an $m \times n$ integer matrix
$\Omega$ of complexity $m(n - 1)$ whose entries are double exponential
in $mn$.
Thus, the binary representation is of an exponential size,
which is one exponent less than the size of
the example from~\cite{KaminskiS21}. Like the example in~\cite{KaminskiS21}, 
this example is also based on the technique developed by Strassen~\cite{Strassen74}.

Namely, we prove the following result.

\begin{theorem}
\label{thm: main}
Let
\[
\Omega
=
\begin{pmatrix}  
\omega_{1,1} & \omega_{1,2} & \cdots & \omega_{1,n}
\\
\omega_{2,1} & \omega_{2,2} & \cdots & \omega_{2,n}
\\
 \vdots & \vdots & & \vdots
\\
\omega_{m,1} & \omega_{m,2} & \cdots & \omega_{m,n}
\end{pmatrix}  
\]
be an integer matrix, with integer entries  
\[
\{\omega_{1,1}, \ldots,  \omega_{m,n}\} = \{ a_1,\ldots,a_N \}  
\]
where $N = mn$,  satisfying  
\begin{align*} 
&N^{N^{N^2}}\ge a_1 >  \frac{1}{2} N^{N^{N^2}}  ,\\
&N^{ \ell N^{N^2 + \ell N-\ell}+N^{N^2} }\ge  a_\ell \ge N^{ \ell N^{N^2 + \ell N-\ell}} , \qquad 
\ell = 2,\ldots,N - 1,\\
&a_N \ge N^{N^{2N^2 - N + 1}}.
\end {align*} 
Then $\cC(\Omega) = m(n - 1)$.
\end{theorem}  

Note that $\max\{ a_1,\ldots,a_N \} = a_N$ and,
for  the choice 
\[
a_N =   N^{N^{2N^2 - N + 1}},  
\]
the entries  
\[\{ \omega_{s,t} :~s = 1,\ldots,m, \  t = 1,\ldots,n \} 
= \{ a_1,\ldots,a_N\}
\] of $\Omega$ are double exponential in $N$.
Note that Theorem~\ref{thm: main} allows us a lot of flexibility in the choice of the 
parameters $a_1,\ldots,a_N$. On the other hand, if one just needs one concrete example
then  Theorem~\ref{thm: expl ai} in Appendix~\ref{app:NonVanishing}   provides an example of a matrix whose entries are slightly smaller than the above values of $a_1,\ldots,a_N$.

This paper is organized as follows.
Section~\ref{sec: compl} consists of two parts.
Section~\ref{sec: linear algorithms}
contains the definition of a linear algorithm and
its associated graph and
in Section~\ref{sec: normalized}, we introduce  {\it normalized\/} linear algorithms  and
state some simple basic complexity results.
The proof of Theorem~\ref{thm: main}
is presented in Section~\ref{sec: proof}. 
We conclude the paper with a short remark
concerning the size of our example.

We also tighten some auxiliary estimates from~\cite{KaminskiS21}, which could be of 
independent interest. For example, see Lemmas~\ref{lem: general height}
and~\ref{lem: nonvanish}; as well as Appendices~\ref{app:Proof} and~\ref{app:NonVanishing} 
for some additional information.

\section{Background from the complexity theory}
\label{sec: compl}

\subsection{Linear algorithms and their associated graphs}
\label{sec: linear algorithms}

A {\em linear algorithm} over a field $\F$ in {\em indeterminates}
$x_1,x_2,\ldots,x_n$ consists of a sequence of operations
$u_i \leftarrow \alpha_i u_{j_i} + \beta_i u_{k_i}$, $i = 1,\ldots,\Lcomp$,
where
\begin{itemize}
\item
$\alpha_i,\beta_i \in \F^\ast$ are the algorithm coefficients; 
\item
$u_i$ is the algorithm {\em variable}
that does not appear in a previous step;
\item
$u_{j_i}$ and $u_{k_i}$ are either indeterminates (namely, belong to the set $\{x_1,\ldots,x_n\}$)  or
the algorithm variables appearing in a previous step 
(that is,
if $u_{j_i}$ and  $u_{k_i}$ are the algorithm variables appearing at step $i$,
then $j_i , k_i < i$).
\end{itemize}

With each algorithm variable $u$ in a linear algorithm
we associate the following linear form $\ell(u)$:
\begin{itemize}
\item
if $u$ is an indeterminate $x_t$,
then $\ell(u)$ is $x_t$; 
\item
if $u$ is the left-hand side of an operation
$u \leftarrow \alpha v + \beta w$,
then $\ell(u)$ is the linear form $\alpha \ell(v) + \beta \ell(w)$.
\end{itemize}

A linear algorithm {\em computes} a linear form $\ell(x_1,\ldots,x_n)$,
if there is a variable, or an indeterminate, $u$ of the algorithm and
a {\em constant} $\gamma \in \F^\ast$
such that 
$\ell(x_1,\ldots,x_n) = \gamma \ell (u)$ 
(thus,
linear algorithms compute linear forms up to scaling by a constant). A 
linear algorithm computes a set linear forms
\[
\cL(x_1,\ldots,x_n) = \{\ell_s(x_1,\ldots,x_n) :~s = 1,\ldots,m \}
\]
if it computes each form $\ell_s(x_1,\ldots,x_n)\in \cL(x_1,\ldots,x_n)$.  

The number $n$ of the variables and the number $m$ of linear forms
is fixed throughout this paper.

\begin{definition}
The complexity $|\cA|$ of a linear algorithm $\cA$ is
the length $\Lcomp$ of its sequence of operations.
\end{definition}   

\begin{definition}
The (additive) complexity of a set of linear forms
is the minimal complexity of a linear algorithm that computes the set.
\end{definition}  

It is known from~\cite{Strassen73} that if a set of linear forms over
an infinite field can be computed in $\Lcomp$ additions
by a {\em straight-line} algorithm (see~\cite[Section~12.2]{AhoHU74}),
then it also can be computed in $\Lcomp$ additions by a linear algorithm.
In other words,
multiplications and divisions ``cannot replace additions.''

With a linear algorithm $\cA$ we associate
a labelled directed acyclic graph $G_\cA=\(V_\cA,E_\cA\)$,
whose set of vertices is
the union of $\{ x_1,\ldots,x_n \}$ and
the set of the variables of $\cA$ and
there is an edge from vertex $v$ to vertex $u$,
if there is an operation of
the form $u \leftarrow \alpha v + \beta w$ or
the form $u \leftarrow \alpha w + \beta v$.
In the former case, the edge is labelled $\alpha$ and,
in the latter case, it is labelled~$\beta$,
see Figure~\ref{fig: graph} below.

\begin{figure}[h]
\begin{picture}(-10,80)(-190,-10)
\put(-240,0){$\bullet$}

\put(-240,-10){$v$}

\put(-160,0){$\bullet$}

\put(-160,-10){$w$}

\put(-200,40){$\bullet$}

\put(-200.4,50){$u (\leftarrow \alpha v + \beta w)$}

\put(-237,3){\vector(1,1){38.1}}

\put(-230,25){$\alpha $}

\put(-157,3){\vector(-1,1){38.1}}

\put(-172,25){$\beta $}
\end{picture}
\caption{Edges and vertices of $G_\cA$.}
\label{fig: graph}
\end{figure}

\par \noindent
We denote the label of edge $e$ by $\lambda(e)$.

\begin{remark}
\label{r: in-degree 2}
By definition,
$| V_\cA | = n + |\cA|$ and the number of vertices of $G_\cA$
of the in-degree $2$ is $|\cA|$.
\end{remark}

Let $ \pi = e_1,\ldots,e_k$ be a path of edges in $G_\cA$.
The {\em weight} $\w(\pi)$ of $\pi$ is defined, recursively,
as follows.
\begin{itemize}
\item
If $\pi$ is of length zero, then $\w(\pi) = 1$; and
\item
$\w(\pi,e) = \w(\pi)\lambda(e)$,
where $\pi,e$ is the path $\pi$ extended with edge $e$.
\end{itemize}

The following correspondence between linear algorithms and
their associated graphs is well-known from the literature, see, for 
example,~\cite[Remark~13.19]{BuergisserCSL97}.

\begin{lemma}
\label{p: sum of weights}
Let
\[
\cA =
\{ u_i \leftarrow \alpha_i u_{j_i} + \beta_i u_{k_i} :~
i = 1,\ldots, | \cA | \}
\]
be a linear algorithm and
let $\Pi_\cA\(x_t,u_i\)$ denote the set of all paths of edges from
the indeterminate $x_t$ to the algorithm variable $u_i$ in $G_\cA$.
Then
\[
\ell(u_i)
=
\sum_{t=1}^n
\left( \sum_{\pi \in \Pi_\cA\(x_t,u_i\)}   \w (\pi) \right)
x_t,
\qquad  
i = 1,\ldots,| \cA |. 
\]
\end{lemma} 

\subsection{Normalized linear algorithms}
\label{sec: normalized}

In this section we introduce a subclass of linear algorithms called
{\em normalized} linear algorithms.
These algorithms have the same computation power,
but are more convenient for dealing with complexity issues.

\begin{definition}
\label{d: normalized}
A linear algorithm is normalized if in each its operation
\[
u_i \leftarrow \alpha_i u_{j_i} + \beta_i u_{k_i}
\]
the coefficient $\alpha_i$ of $u_{j_i}$ is $1$.
The coefficient $\beta_i$ of $u_{j_k}$, that also may be $1$, is called
a {\it proper}  coefficient.
\end{definition}

We say that a label is {\it proper} if it is 
a proper coefficient  of the algorithm. 

The result below immediately follows from Definition~\ref{d: normalized}, 
the definition of the associated graph $G_\cA$ of an algorithm $\cA$ and Remark~\ref{r: in-degree 2}.

\begin{lemma}
\label{p: labels}
The additive complexity of a normalized linear algorithm $\cA$ equals  
the number of proper labels
of its associated graph $G_\cA$.
\end{lemma} 

Furthermore, we also have the following result, given in~\cite[Proposition~6]{KaminskiS21}.

\begin{lemma}
\label{p: normalized}
For each linear algorithm there is a normalized linear algorithm of
the same complexity that computes the same set of linear forms.
\end{lemma}

From now on, by Lemma~\ref{p: normalized},
we assume that all linear algorithms under consideration are normalized.

\section{Proof of Theorem~\ref{thm: main}}
\label{sec: proof}

\subsection{Outline} 
The proof  is based on 
\begin{itemize}
\item 
an explicit form of the {\it Perron algebraic dependence theorem\/}, see~\cite{Perron27}, 
see Lemma~\ref{lem: general height}; 

\item a relationship between complexity of linear forms and zeros of some multivariate 
polynomials, see  Lemma~\ref{lem: transcendence}; 

\item a new construction of a reasonably small integer vector which provides a
common {\it nonzero\/} to a large family of multivariate polynomials, see 
Lemma~\ref{lem: nonvanish}.
\end{itemize}

\subsection{Annihilating polynomials} 

To formulate a fully explicit form   of the Perron  theorem~\cite{Perron27} we introduce 
the following definition.   

\begin{definition}
We say that $P(Z_1,\ldots,Z_N) \in \Z[Z_1,\ldots,Z_{N}]$
is an {\it annihilating polynomial\/}  of 
$P_\k\(X_1,\ldots,X_{N-1}\) \in \Z[X_1,\ldots,X_{N-1}]$,
$k =1, \ldots, N$,
if $P$ is a nonzero polynomial and 
\[
P\(P_1\(X_1,\ldots,X_{N-1}\),\ldots, P_N\(X_1,\ldots,X_{N-1}\)\) = 0. 
\]
\end{definition}

We start with the following result, that is essentially due to Perron~\cite[Theorem  57, p.~129]{Perron27},
(see also~\cite[Theorem~1.1]{Ploski05} for a self-contained proof). 

\begin{lemma}
\label{lem: dependence}
Let
\[P_\k(X_1,\ldots,X_{N-1}) \in \Z[X_1,\ldots,X_{N-1}]
\]
with $\deg P_\k(X_1,\ldots,X_{N-1}) = d_\k$, $\k = 1,\ldots,N$.
Then there exists an annihilating polynomial
$P(Z_1,\ldots,Z_N) \in \Z[Z_1,\ldots,Z_N]$ of $P_1,\ldots,P_N$
such that
\[
\deg P
\leq
\frac{d_1 \times \cdots \times d_N} { \min \{ d_1,\ldots,d_N \}}.
\]
\end{lemma}

We also use $H(P)$ for the {\it naive height\/} of  a  polynomial $P$ over $\C$ (in one or several variables), that is,  the largest absolute value of its coefficients.  

Lemma~\ref{lem: general height} below is a slight improvement of~\cite[Proposition~23]{KaminskiS21}.

\begin{lemma}
\label{lem: general height}
Let $P(Z_1,\ldots,Z_N) \in \Z[Z_1,\ldots,Z_N]$ be
an annihilating polynomial of 
\[
P_\k(X_1,\ldots,X_{N-1}) \in \Z[X_1,\ldots,X_{N-1}],
\qquad k = 1,\ldots,N. 
\]
There exists another annihilating polynomial 
\[
Q(Z_1,\ldots,Z_N) \in \Z[Z_1,\ldots,Z_N]
\] 
of $P_1,\ldots,P_N$
of degree and height
\begin{align*}
& \deg Q \le \deg P,\\
& H(Q) \le \({\binom{\deg P + N}{N}}^{1/2}  N^{\dm \deg P}
\Hm^{\deg P} \)^{\binom{\deg P + N}{N} - 1},
\end{align*}  
 respectively, where
\begin{align*}
& \dm = \max \{ \deg P_\k :~\k = 1,\ldots, N \}, \\
& \Hm = \max \{ H(P_\k) :~\k = 1,\ldots, N \}.
\end{align*} 
\end{lemma}   

We present a proof of Lemma~\ref{lem: general height} 
in Appendix~\ref{app:Proof}. 

\subsection{Complexity of linear forms and vanishing of polynomials}

The first step in the proof of Theorem~\ref{thm: main}
is similar to that in~\cite{KaminskiS21}. 
It also resembles some previous results of this type, 
see, for example,~\cite[Lemma~9.28]{BuergisserCSL97} 
or~\cite[Lemma~2.3]{Strassen74}, however, 
it seems to be new, see also Appendix~\ref{app:NonVanishing}.

\begin{lemma}
\label{lem: transcendence}
 If $\cC(\Delta) < m(n - 1)$,
then for some nonzero polynomial with integer coefficients
$Q\(Z_{1,1},\ldots,Z_{m,n}\)$ of degree and height 
\begin{align*}
& \deg Q \le N^{N-1},\\
& H(Q)  \le \frac{1}{2} N^{N^{N^2}} 
\end{align*}     
 respectively,
where $N = mn$, we have
$Q\(\delta_{1,1},\ldots,\delta_{m,n}\) = 0$.
\end{lemma}

\begin{proof}
The   assumption  $\cC(\Delta) < m(n - 1)$ 
implies that $N = mn    > \cC(\Delta) + m \ge m \ge 1$. Thus $N\ge 2$. 
If for some $i = 1,\ldots,m$, $j = 1,\ldots,n$ we have $\delta_{i,j} = 0$,
then we take
\[
Q(Z_{1,1},\dots,Z_{m,n}) = Z_{i,j}
\]
in which case $\deg Q=H(Q)=1$.
So, assume now that $\Delta$ has no zero entries. However in this case 
 $\cC(\Delta) > 0$ and   we infer that $N = mn > \cC(\Delta) + m \ge 2$. 
 
 Thus we can now assume 
that $N \ge 3$.

Recall that we represent a linear form
$\ell(x_1,\ldots,x_n) = \delta_1 x_1+ \ldots +  \delta_n x_n$
by the product $(\delta_1,\ldots,\delta_n) \bx$,
where $\bx$ is the (column) vector of the indeterminates $x_1,\ldots,x_n$
as in~\eqref{eq: vector}. 
Similarly,
we represent a set of linear forms
\[
\ell_s (x_1,\ldots,x_n) = \sum_{t=1}^n \delta_{s,t} x_t \, ,
\qquad 
s = 1,\ldots,m, 
\]
by a matrix-vector product $\Delta \bx$,
where the $s$th row of the matrix $\Delta$ is the row vector
$(\delta_{s,1},\ldots,\delta_{s,n})$ of the coefficients of
$\ell_s (x_1,\ldots,x_n)$, see~\eqref{eq: sets}.

Let $\cA$ be a linear algorithm that computes~\eqref{eq: sets} and
let $G_\cA = (V_\cA,E_\cA)$ be its associated labelled graph.

Let $u_{i_s}$ and $\gamma_s$, $s = 1,\ldots,m$,
be the algorithm variables and the respective constants such that
\[
\ell_s(x_1,\ldots,x_n)
=
\sum_{t=1}^n \delta_{s,t} x_t
=
\gamma_s \ell(u_{i_s}).
\]

Then, by Lemma~\ref{p: sum of weights}, we have
\begin{equation}
\label{eq: delta_st}
\delta_{s,t}
=
\gamma_s
\sum_{\pi \in \Pi_\cA\(x_t,u_{i_s}\)}  \w (\pi)
=
P_{s,t}\(\gamma_s,\beta_1,\ldots,\beta_{|\cA|}\)
\end{equation}
for some polynomials $P_{s,t}\(Y_s,X_1,\ldots,X_{|\cA|}\)$ in $|\cA|+1$ variables,
$s = 1,\ldots,m$ and $t = 1,\ldots,n$, where $\beta_1,\ldots,\beta_{|\cA|}$ are 
the graph labels as in  Definition~\ref{d: normalized}.
It follows from Lemma~\ref{p: sum of weights} that
$\deg P_{s,t} \leq N$ and $H(P_{s,t}) = 1$,
$s = 1,\ldots,m$ and $t = 1,\ldots,n$.  

If the number $|\cA|$ of the proper labels is less than $m(n - 1)$,
then the total number of $\beta$ and $\gamma$ variables is less than $mn$  --- see Lemma~\ref{p: labels},
implying that these $mn$ polynomials are algebraically dependent. 
 
Let $P(Z_{1,1},\ldots,Z_{m,n})$ be the  annihilating polynomial with integer coefficients
which is provided by Lemma~\ref{lem: dependence}.

Now Lemma~\ref{lem: general height} gives a polynomial $Q$ with
$$
\deg Q \le \deg P \leq N^{N-1}.
$$

It is useful to observe that the assumption $\cC(\Delta) < m(n - 1)$ implies then
inequality $1 <  m(n - 1)$, which in turn yields  $N = mn\ge 3$. 

Note that for $N \geq 3$ we have
$$
\( 1 + \frac{1}{N^{N-2}} \)^N
<\( 1 + \frac{1}{N} \)^N < e<3\le \frac{N !}{2}. 
$$
Hence,  
 \begin{align*}
\binom{\deg P + N}{N} & < \frac{\(\deg P + N\)^N}{N!}  \le\frac{\(N^{N-1}+ N\)^N}{N!}\\
& =  N^{N(N-1)}\( 1 + \frac{1}{N^{N-2}} \)^N\frac{1} {N!}  < 
\frac{1}{2} N^{N(N-1)}
\end{align*}  
and applying Lemma~\ref{lem: general height} with $\Hm =1$ and $\dm \le N$, we obtain 
 $$
 \binom{\deg P + N}{N}^{1/2}N^{\dm \deg P }< N^{N^2/2+N^{N}} \le  N^{2N^{N}}, 
 $$
 and then we see that $H(P) <N^\kappa$ with
 $$
\kappa =2N^{N} \binom{\deg P + N}{N} < N^{N} N^{N^2-N} = N^{N^2}. 
 $$
Thus, since the inequality is strict,  we obtain
 $$\kappa \le  N^{N^2}-1.
 $$

Hence
 $$
H(Q)  < N^{N^{N^2}-1}  \le \frac{1}{2} N^{N^{N^2}} 
$$ 
for $N \ge 3$ 
(note that this is a slight improvement of~\cite[Corollary~24]{KaminskiS21}).

It follows from~\eqref{eq: delta_st}, and because $Q$ is  an annihilating polynomial, that
\[
Q\(P_{1,1}\(\gamma_1,\beta_1,\ldots,\beta_{|\cA|}\)
,\ldots,
P_{m,n}\(\gamma_m,\beta_1,\ldots,\beta_{|\cA|}\)\)
=
0, 
\]
implying
\[
Q\(\delta_{1,1},\ldots,\delta_{m,n}\) = 0,
\]
which concludes the proof. 
\end{proof}

\begin{remark}
\label{r: labels}
Note that the polynomials $P_{s,t}$ constructed in the proof of  Lemma~\ref{lem: transcendence} do not depend on the graph labels $\beta_1,\ldots,\beta_{|\cA|}$.
\end{remark}

\subsection{Zeros and nonzeros of polynomials}

The following bound on zeros of polynomials is very well~known, 
see, for example,~\cite[Theorem~4.2]{Mig}. 

\begin{lemma}
\label{lem: root bound}
 Let $f\in\Z[X]$ be a nonzero polynomial  and let $\alpha$ be a complex root of $f$. Then 
 $|\alpha|< H(f)+1$. 
\end{lemma}
 
 We now establish our main technical tool Lemma~\ref{lem: nonvanish}. In Appendix~\ref{app:NonVanishing},
 we compare this result with several statements of a similar flavour about non-vanishing of 
 polynomials.
 
\begin{lemma}
\label{lem: nonvanish}
Let $P(Z_1,\ldots,Z_N)$ be a nonzero polynomial with integer coefficients
of degree at most $d\ge 3$ and height $H$.
Then, for integers 
\[a_1 > H \mand  a_\ell \geq 2H a_{\ell - 1}^d,\quad 
\ell = 2,\ldots,N,
\] 
we have $P\(a_1,\ldots,a_N\)\ne 0$.
\end{lemma}

\begin{proof}
 The proof is by induction on $N$. 
 The case  $N=1$ plainly follows from   Lemma~\ref{lem: root bound}, since $a_1\ge H+1$.  
 Assume  that $N\ge 2$ and that the result holds for $N-1$. 

Write 
 \[
 P(X_1,\ldots,X_N)=\sum_{j=0}^d P_j(X_1,\ldots,X_{N-1}) X_N^j.
 \]
The polynomials $P_j(X_1,\ldots,X_{N-1})\in\Z[X_1,\ldots,X_{N-1}]$ have degree at most  $d$ and height
at most  $H$ and one of them is not zero. From the induction hypothesis, we deduce that the integers $P_j(a_1,\ldots,a_{N-1}) $ with $0\le j\le d$ are not all zero. Hence   the polynomial 
 \[
 F (X)=P(a_1,\ldots,a_{N-1},X)=\sum_{j=0}^d P_j(a_1,\ldots,a_{N-1}) X^j\in\Z[X]
 \]
 is not zero. Its degree is at most $d$. We claim that the height of $F$ is strictly less than
$2H a_{N-1}^d$.

 Let us write, for $j=0,1,\ldots,d$,  
 \[
 P_j(X_1,\ldots,X_{N-1})=\sum_{j_1+\cdots+j_{N-1}\le d-j}p_{j_1,\ldots,j_{N-1},j}X_1^{j_1}\cdots X_{N-1}^{j_{N-1}}
\]
with $p_{j_1,\ldots,j_{N-1},j}\in\Z$, $|p_{j_1,\ldots,j_{N-1},j}|\le H$. 
We have  
 \begin{align*}
 |P_j(a_1,\ldots,a_{N-1}) | 
 &\le
 \sum_{j_1+\cdots+j_{N-1}\le d-j}|p_{j_1,\ldots,j_{N-1},j}| a_1^{j_1}\cdots a_{N-1}^{j_{N-1}}
 \\
 &
 \le H \sum_{j_1+\cdots+j_{N-1}\le d-j} a_1^{j_1}\cdots a_{N-1}^{j_{N-1}}
 \\
 &
 \le H  \sum_{\ell=0}^{d-j}(a_1+a_2+\cdots +a_{N-1})^\ell.
 \end{align*}
Denote 
\[S =a_1+a_2+\cdots +a_{N-1}.
\]
In the case $N=2$, we have $S=a_1\ge 2$ and 
\[
   \sum_{\ell=0}^d S^\ell =\frac {a_1^{d+1}-1} {a_1-1} < 2 a_1^d,
 \]
 which proves the claim on the height of $F$ in this case. 
 
 Assume now $N\ge 3$. 
Since $H\ge 1$,  we have $a_1\ge 2$ and $a_2\ge 8$, implying $S \ge 10$. Hence 
\[
\frac S {S-1}\le \frac {10}{9}
\] 
and
\begin{equation}
\label{eq: sum Dl}
  \sum_{\ell=0}^d S^\ell <\frac {S} {S-1} S^d\le  \frac {10}{9}  S^d.
\end{equation}

From $a_1\ge 2$ and from $a_\ell \ge 2 a_{\ell-1}^d$, 
by a straightforward induction we conclude
that 
\[a_\ell  \ge 2^{d^{\ell-1}}, \qquad 
\ell = 1,\ldots,N. \]
Using that  $2^k \ge 2k$ for any integer  $k \ge 0$, we derive
\[
  a_{N-2}^{d-1}  \ge  2^{(d-1) d^{N-3}}\ge  2 (d-1)d^{N-3}  
 > \frac{5}{4}d^{N-2}  \ge \frac{5}{4} d(N-2).
\]  
  As a consequence, we have 
\[
  a_1+\cdots+a_{N-2} < (N-2) a_{N-2}\le \frac { (N-2)a_{N-1}}{2Ha_{N-2}^{d-1}}\le
  \frac  {2} {5d}  a_{N-1},
  \]
  hence
\[  S< \left( 1+\frac 2 {5d}\right)a_{N-1}
  \le  \left(\frac 3 2 
 \right)^{1/d} a_{N-1}.
 \]
Recalling~\eqref{eq: sum Dl}, we see that 
\[
  \sum_{\ell=0}^d S^\ell <  \frac {10}{9} S^d \le \frac {5}{3} a_{N-1}^d. 
  \]
Hence we obtain the desired claim that  the height of $F$ is at most  
 \[
 \frac {5}{3} H a_{N-1}^d<
 2Ha_{N-1}^d. 
 \]

 Using the assumption $a_N\ge 2Ha_{N-1}^d$
together with  Lemma~\ref{lem: root bound}, we conclude $F(a_N)\not=0$ and the result follows.  
\end{proof}

 \subsection{Concluding the proof of Theorem~\ref{thm: main}} We can assume that 
 $\cC(\Omega) < m(n - 1)$ as otherwise there is nothing to prove. 
 As we have mentioned in the proof of Lemma~\ref{lem: transcendence}, we always 
 have $N = mn \ge 3$.
Then we just check that   the conditions of Lemma~\ref{lem: nonvanish}  are fulfilled if one selects the integers $a_1,\ldots,a_N$ satisfying 
 \begin{align*} 
& 2H\ge a_1 >H,  \\
& (2H)^{\ell d^\ell +1} \ge a_\ell  > (2H)^{\ell d^\ell}  , \qquad 
\ell = 2,\ldots,N - 1,\\
&a_N \ge  (2H)^{Nd^N}. 
\end {align*} 
 Indeed, these inequalities yield for $d \ge N$ 
 \[
 a_2\ge (2H)^{2N^2} \ge  (2H)^{N+1}\ge 2Ha_1^N 
\]
and, for $\ell=2,3,\ldots,N-1$,
\[
 a_{\ell+1} \ge (2H)^{(\ell +1)d^{\ell +1} } \ge (2H)^{\ell d^{\ell+1} +d}
 \ge 2Ha_\ell^d. 
 \]
Taking 
\[d = N^{N-1} \mand H = \frac{1}{2} N^{N^{N^2}} 
\]
as  for the polynomial $Q$ of  Lemma~\ref{lem: transcendence}, after simple calculations, 
we see that the conclusions of  Lemmas~\ref{lem: transcendence} and~\ref{lem: nonvanish} about vanishing of $Q\(a_1,\ldots,a_N\)$ contradict to each other. Hence our initial assumption  $\cC(\Omega) < m(n - 1)$ has been false, which 
concludes the proof.

\section{Concluding remark}

Even though, if the degree and the height
of the ``annihilating'' polynomial $P$ from the proof
of Lemma~\ref{lem: transcendence} can be reduced,  in our approach, 
the entries of $\Omega$   in Theorem~\ref{thm: main} remain double exponential in $N$. 
Of course it would be interesting to find an explicit example that is only
exponential  in the matrix size, whose existence is 
shown in Theorem~\ref{thm: non-eff} below.
However, it is to be expected that this challenge requires a different approach.

\appendix

\section{Proof of Lemma~\ref{lem: general height}}\label{app:Proof}

We need the following two well known statements

The first one is a bound on  the height of the product of polynomials over $\Z$, which can be
found, for example,  in~\cite[Lemma~1.2(1)(b) p.~531]{KrickPS01}.

\begin{lemma} 
\label{lem: height of the product}
Let $P_\k \in \Z[X_1,\ldots,X_{N-1}]$, $\k = 1,\ldots,\ell$ and 
$$R = \prod_{\k=1}^\ell P_k.
$$
Then 
\[
H (R)
\leq
N^{ \deg R}
\prod_{\k=1}^\ell H(P_\k).
\]
\end{lemma} 

Instead of using the classical {\it Siegel lemma\/}  used in~\cite[Lemma 22]{KaminskiS21}, here we 
rely on the following statement from linear algebra.

\begin{lemma}
\label{lem: Siegel}
Let $b_{i,j}$, $i=1,\ldots,I$, $j=1, \ldots, J $, be rational integers, not all of which are $0$. Let 
\[
B=\max\{|b_{i,j}|:~i=1, \ldots, I, \ j=1, \ldots, J\} .
\]
Assume that there exists a nonzero solution $(z_1, \ldots, z_I)$ to the system of  $J$ linear homogeneous equations in $I$ variables
\[
\sum_{i=1}^I b_{i,j} z_i = 0, \qquad j=1, \ldots, J.
\]
Then this system has
a nonzero integer solution $\bv=(v_1, \ldots, v_I)$ with 
\[
\max_{1\le i\le I} |v_i| \le  \(\sqrt{I-1}B\)^{I-1}.
\] 
\end{lemma} 

\begin{proof} 
Let $L$ be the rank of the matrix $ \(b_{i,j}\)_{i,j=1}^{I,J} \in \Z^{I\times J}$. The assumption that the system has a nonzero solution implies $L<I$. We consider a nonsingular  $L\times L$ minor $D$ of this matrix; without loss of generality we may assume that it is 
$ \(b_{i,j}\)_{i,j=1}^{L} \in \Z^{L\times L}$. 
Let $(u_1,\ldots,u_L)\in\Q^L$ be the solution of the non homogeneous linear system of $L$ equations in $L$ unknowns 
\[
\sum_{i=1}^L b_{i,j} u_i = -b_{L+1,j}, \qquad j=1, \ldots, L, 
\]
 provided by Cramer's rule.
Set $v_i=u_ i \det D$ for $1\le i\le L$, $v_{L+1}=- \det D$ and $v_i=0$ for $L+1<i\le I$. Then $\bv$ is a nonzero integer solution to the system of  $J$ linear homogeneous equations in $I$ variables. From Hadamard's upper bound for a determinant~\cite{Had},  we deduce 
\[
\max_{1\le i\le I} |v_i|\le (\sqrt L B)^L, 
\]
which concludes the proof.
\end{proof}

We are now ready to proceed with the proof of Lemma~\ref{lem: general height}. 

We employ the following notation:
\begin{itemize}
\item $\bX = (X_1,\ldots,X_{N-1})$ and
$\bZ = (Z_1,\ldots,Z_N)$ are vectors of variables;
\item
$\bi = (i_1,\ldots,i_N)$ and $\bj = (j_1,\ldots,j_{N-1})$
are vectors of non-negative integers; 
\item
$\bX^{\jscript} = \prod_{s=1}^{N-1} X_s^{j_s}$ and 
$\bZ^{\iscript} = \prod_{k=1}^N Z_k^{i_k}$
are multivariate monomials.
\end{itemize}

We search for an annihilating polynomial $Q$ of $P_1,\ldots,P_N$ in the form 
\begin{equation}
\label{eq: Q}
Q(Z_1,\ldots,Z_N)
=
\sum_{\iscript:\, i_1+\ldots+i_N \leq \dgP} v_{\iscript} \bZ^{\iscript},
\end{equation}
with unknown coefficients $v_{\iscript}$ to be determined. 

To find the coefficients $v_{\iscript}$\, of $Q(Z_1,\ldots,Z_N)$,
we substitute  the polynomials $P_\k(X_1,\ldots$, $X_{N-1})$ for $Z_\k$,
$\k = 1,\ldots,N$, in~\eqref{eq: Q}, obtaining
\begin{align*} 
& Q\(P_1(X_1,\ldots,X_{N-1}),\ldots, P_N(X_1,\ldots,X_{N-1})\)\\
&
\qquad \qquad \quad  
=
\sum_{\iscript:\, i_1+\ldots+i_N \leq \dgP} v_{\iscript}
\prod_{\k=1}^N
P_\k^{i_\k}(X_1,\ldots,X_{N-1})
=
0.
\end{align*}

Let
\[
\prod_{\k=1}^N
P_\k^{i_\k}(X_1,\ldots,X_{N-1})
=
\sum_{\jscript:\, j_1+\ldots + j_{N-1}
\le
\dm \dgP } \bc_{\iscript,\jscript} \bX^{\jscript}. 
\]
Since
\begin{align*}
& \sum_{\iscript:\, i_1+\ldots+i_N \leq \dgP }  v_{\iscript}
\sum_{\jscript:\, j_1+\ldots + j_{N-1}
\le
\dm \dgP }\bc_{\iscript,\jscript} \bX^{\jscript}\\
&\qquad \qquad  =
\sum_{\jscript:\, j_1+\ldots + j_{N-1} \le \dm \dgP }
\bX^{\jscript}
\left( \sum_{\iscript:\, i_1+\ldots+i_N \leq \dgP }
\bc_{\iscript,\jscript} v_{\iscript} \right)
=
0, 
\end{align*} 
 we obtain a system of linear homogeneous equations
\begin{equation}
\label{eq: Syst cij} 
\sum_{\iscript:\, i_1+\ldots+i_N \leq \dgP }
\bc_{\iscript,\jscript} \, v_{\iscript} = 0,
\qquad \bj: \ j_1+\ldots + j_{N-1} \le \dm \dgP 
\end{equation}  
in 
\begin{equation}
\label{eq: def I} 
I = \binom {\dgP+N}{N}
\end{equation}
unknowns $v_{\iscript}$ (the coefficients of $Q(Z_1,\ldots,Z_N)$).

We also note that for the coefficients $\bc_{\iscript,\jscript}$
of the system of 
linear equations~\eqref{eq: Syst cij} we have
\[
\max_{\iscript,\jscript} |\bc_{\iscript,\jscript}|
\le
\max _{i_1+\ldots+i_N \leq \dgP} H\(\prod_{\k=1}^N P_\k^{i_\k}\), 
\]
where $\bi$ and $\bj$ run through the vectors
with $i_1+\ldots+i_N \leq \dgP$ 
and $j_1+\ldots + j_{N-1} \le \dm \dgP $, respectively.

Hence, by Lemma~\ref{lem: height of the product}
with 
$$
R = \prod_{k=1}^N P_k^{i_k}
$$
we have 
\begin{equation}
\label{eq: h hat}
\max_{\iscript,\jscript} |\bc_{\iscript,\jscript}| 
\leq N^{\dm \dgP}  \Hm^{\dgP}, 
\end{equation}
because
$$
\ell = i_1 + \cdots + i_n \le \deg P .
$$

Since, by our assumption on the polynomial $P$,
this system has a nonzero solution,  
we can apply Lemma~\ref{lem: Siegel} with the bound~\eqref{eq: h hat}, obtaining
that~\eqref{eq: Syst cij} 
has a solution with 
\begin{align*} 
\max\{ v_{\iscript}:~\bi = (i_1, \ldots, i_N)\   \text{with} \ & 
i_1 +\ldots+i_N \leq \dgP\}\\
& \le  \(\sqrt{I-1} N^{\dm \dgP} \Hm^{\dgP} \) ^{I-1}, 
\end{align*} 
where $I$ is given by~\eqref{eq: def I}, 
which concludes the proof.

\section{Non-vanishing of polynomials}\label{app:NonVanishing}

Our main technical tool, namely Lemma~\ref{lem: nonvanish}, is a non-vanishing result on polynomials. Similar results can also be found in~\cite[Lemma~9.30]{BuergisserCSL97} and~\cite{Fuk}. 
In~\cite{Fuk}, Fukshansky's goal is to prove the existence of a point outside a hypersurface, while our aim it to give sufficient conditions for a point to satisfy such a condition. 

Furthermore,  the assumption of~\cite[Lemma~9.30]{BuergisserCSL97} that the height is at most $3$ is too restrictive for our purpose. Another variant of our Lemma~\ref{lem: nonvanish}
 is~\cite[Lemma~2]{Koi}, but this statement needs to be corrected since a counterexample is   $p=2$, $H>d$, 
\[
P(X_1,X_2)=HX_1^d-X_2, \quad a_1=H+1,\quad a_2=H(H+1)^d,
\]
a corrected version has been given in~\cite{Koi-Corr}\footnote{After the paper appeared online in {\it Comp. Compl.\/}, Pascal Koiran informed us that he had corrected~\cite[Lemma~2]{Koi}  in~\cite{Koi-Corr}. Unfortunately, it was too late to update the journal version.}.
According to Koiran, his~\cite[Lemma~2]{Koi} 
{\em is essentially due to Heintz and Schnorr}~\cite[Lemma~4.2]{HS1980}, 
who in turn, attribute  it  to Kronecker~\cite{Kro}.

We  now give the following variant of~\cite[Lemma~4.2]{HS1980}  and  Lemma~\ref{lem: nonvanish}.
 
\begin{lemma} 
\label{lem:Kron} 
Let $P\in\Z[Z_1,\ldots,Z_N]$ be a nonzero polynomial of partial degree less than  $D$ in each variable and height $H$. Then for any 
complex number $\xi$ with  $|\xi|\ge H+1$ we have 
\[
P\(\xi,\xi^D,\xi^{D^2},\ldots,\xi^{D^{N-1}}\)\not=0.
\]
\end{lemma} 

\begin{proof}
We use the so-called Kronecker substitution~\cite[Pages~11--12]{Kro} by considering the univariate polynomial 
\[
R(T)=
P(T,T^D,T^{D^2},\ldots,T^{D^{N-1}}).
\]
This is not the zero polynomial. Indeed, if 
\[
P(Z_1,\ldots,Z_N)=\sum_{j_1=0}^{D-1}\cdots \sum_{j_N=0}^{D-1}a_{j_1,\ldots,j_N}Z_1^{j_1}\cdots Z_N^{j_N},
\]
then 
\[
R(T)=\sum_{k=0}^{D^N-1} b_k T^k,
\quad
\]
where $b_k=a_{j_1,\ldots,j_N}$ and  the expansion in basis $D$ of $k$ is 
\[
k=j_1+j_2D+\cdots+j_ND^{N-1}.
\]
Since one at least of the $a_{j_1,\ldots,j_N}$ is not zero, one deduces that  at least one of the coefficient  $b_k$, 
$ k  = 0, \ldots,  D^N-1$,  is not zero. Since the height of $R$ is the same as the height of $P$, we may use Lemma~\ref{lem: root bound} and conclude the proof. 
\end{proof}

Combining Lemma~\ref{lem:Kron}  with Lemma~\ref{lem: transcendence} now immediately derive a slightly less flexible version of Theorem~\ref{thm: main}.

\begin{theorem}
\label{thm: expl ai}
Let
\[
\Omega
=
\begin{pmatrix}  
\omega_{1,1} & \omega_{1,2} & \cdots & \omega_{1,n}
\\
\omega_{2,1} & \omega_{2,2} & \cdots & \omega_{2,n}
\\
 \vdots & \vdots & & \vdots
\\
\omega_{m,1} & \omega_{m,2} & \cdots & \omega_{m,n}
\end{pmatrix}  
\]
be an integer matrix,  with integer entries  
\[
\{\omega_{1,1}, \ldots,  \omega_{m,n}\} = \{ a_1,\ldots,a_N \}  
\]
where $N = mn$,  satisfying 
$$
a_i  = N^{N^{N^2 + (i-1) (N -1)}}, \qquad i = 1, \ldots N.
$$ 
Then $\cC(\Omega) = m(n - 1)$.
\end{theorem}

\section{Existence of simply exponential bounds}
\label{sec:count} 

We now show by a counting argument the existence of  matrices $\Omega$ with integer entries of single exponential size, 
which also achieve the largest possible complexity $\cC(\Omega) = m(n - 1)$.

We start with a slight improvement of a well know result
which asserts that  a polynomial $F(X_1, \ldots, X_n)$ in $n$ variables and  of partial degree at most  $D$, 
has a non-zero
 in the box $[0,D]^n$, see~\cite[Lemma~2.1]{Fuk}, \cite[Corollary~1]{Schwartz80} and  \cite[Theorem~1]{Zippel79}. 
Since this result, also called sometimes  the {\it Zippel Lemma\/}, has been used in many applications 
we believe that  our  version  is of independent  interest.

\begin{lemma}
  \label{lem:TotalDegree}
  Let $P\in\C[X_1,\dots,X_n]$ be a nonzero polynomial of total degree  at most $D$.  Then  at least 
  one of the numbers $P(a_1,\dots,a_n)$, where  $a_1,\dots,a_n$ are nonnegative integers with  $a_1+\cdots+a_n\le D$, is nonzero. 
  \end{lemma}

  \begin{proof}
 We proceed by induction on $n$. For $n=1$ the result is clear. 
 
 Assume that $n\ge 2$ and that the result holds for a polynomial in $n-1$ variables. Now we proceed by induction on $D$. For $D=0$ the polynomial $P$ is constant with $P(0)\neq 0$ and the result is true.
 
Assume that $D\ge 1$ and that the result holds for $D-1$. Assume $P\in\C[X_1,\ldots,X_n]$ has $P(a_1,\dots,a_n)=0$, for all $a_1,\ldots,a_n$ nonnegative integers with $a_1+\cdots+a_n\le D$. Then the polynomial $P(X_1,\dots,X_{n-1},0)$, 
has degree at most  $\D$ and 
  vanishes at all points  $(a_1,\ldots,a_{n-1})$ with  nonnegative components 
  satisfying  $a_1+\cdots+a_{n-1}\le D$. By the induction hypothesis for $n-1$, we deduce $P(X_1,\dots,X_{n-1},0)=0$. 
Therefore, there exists a polynomial $\widetilde P\in\C[X_1,\dots,X_n]$ such that $P(X_1,\ldots,X_n)=X_n\widetilde P (X_1,\dots,X_n)$. The polynomial $\widetilde P$ is of degree at most   $\D-1$ and  vanishes at all integer points $(a_1,a_2,\dots,a_n)$ with  $a_i\ge 0$, $i=1,\dots,n-1$,  and $a_n\ge 1$, $a_1+\cdots+a_n\le D$. We now use the induction hypothesis for $D-1$ for the polynomial $\widetilde P(X_1,\dots,X_{n-1},X_n+1)$ to deduce $\widetilde P=0$, hence $P=0$. which contradicts our assumption that $P$ is a nonzero polynomial. 
\end{proof}

Clearly, Lemma~\ref{lem:TotalDegree} implies that the matrix 
  $$
  \left(
  a_1^{i_1}\cdots a_n^{i_n}
  \right)_{\substack{{a_1+\cdots+a_n\le D} \\
  {i_1+\cdots+i_n\le D}}},
  $$
 (where $a^i=1$ for $a=i=0$)
  is non-singular.

\begin{theorem}
\label{thm: non-eff}
Given $n$ and $m$, there exists an integer matrix 
\[
\Omega
=
\begin{pmatrix}  
\omega_{1,1} & \omega_{1,2} & \cdots & \omega_{1,n}
\\
\omega_{2,1} & \omega_{2,2} & \cdots & \omega_{2,n}
\\
 \vdots & \vdots & & \vdots
\\
\omega_{m,1} & \omega_{m,2} & \cdots & \omega_{m,n}
\end{pmatrix}  
\]
with nonnegative integer entries  
\[\{ \omega_{s,t} :~s = 1,\ldots,m, \  t = 1,\ldots,n \}, 
\] 
where  $N = mn$,   satisfying
\[
 \sum_{s=1}^m\sum_{t=1}^n  \omega_{s,t}  \le   2^{2N} N^{3N-1} , 
\]
such that $\cC(\Omega) = m(n - 1)$.  
\end{theorem}

\begin{proof} We first recall that the polynomial $P$ 
of  Lemma~\ref{lem: transcendence} depends only the 
graph associated with the  corresponding normalized linear algorithm $\cA$, 
 see Remark~\ref{r: labels}. 

We now denote by   $G(C,n)$ the number of graphs associated with all possible  normalized linear algorithms of complexity $C$ in indeterminates $x_1,\ldots,x_n$. 

We estimate  $G(C,n)$ using an inductive  argument which 
is similar to that used in the proof of~\cite[Proposition~34]{KaminskiS21}. 

Clearly we have
$$
G(1,n) =  n(n+1)/2
$$
which corresponds to all possible choices of  $i,j \in \{1, \ldots, n\}$ in a linear operation
of the form
\[
u_1 \leftarrow x_i+ \beta_1 x_j.
\]

Next, 
we have 
$$
G(C+1,n) \le G(C,n) (C+n) (C+n+1)/2, 
$$
where the factor  $(C+n) (C+n+1)/2 $ comes from all possible choices of
\[
v,w \in \{ x_1,\ldots,x_n \} \cup \{ u_1,\ldots,u_C \}
\]
in the last linear operation $u_{C+1} = v + \beta_{C+1} w$ of the algorithm.

Using that $(C+n) (C+n+1)/2\le (C+n)^2$ for $C \ge 0$ and $n \ge 1$ we derive
$G(C+1,n) \le G(C,n) (C+n)^2$ and thus we obtain 
\begin{equation}
\label{eq: bound G}
G(C,n) \le (C+n)^{2C}. 
\end{equation}

We now set 
$$
G_0 = G(m(n-1) -1,n) . 
$$

Multiplying all $G_0$ polynomials $P$ corresponding to such distinct 
graph associated with a normalized linear algorithm of complexity  $m(n-1) -1$, 
we obtain a polynomial $Q$ of degree 
$$
 \deg Q \le G_0 N^{N-1}. 
 $$ 
 Recalling~\eqref {eq: bound G}, we obtain 
 \begin{align*}
 \deg Q & \le (m(n-1) -1+n)^{2m(n-1) -2} N^{N-1}\\
 & \le (2N)^{2N} N^{N-1}
 \le 2^{2N} N^{3N-1} .
 \end{align*}
 Invoking Lemma~\ref{lem:TotalDegree}  we conclude the proof.
 \end{proof}

\bibliographystyle{plain}

\end{document}